\newcommand{\al}{\mbox{$\alpha $}}
\newcommand{\alb}{\mbox{$\bar{\alpha }$}}
\newcommand{\s}{\mbox{$\sigma $}}
\newcommand{\xb}{\mbox{$\bar{x}$}}
\newcommand{\zb}{\mbox{$\bar{z}$}}
\newcommand{\tb}{\mbox{$\bar{t}$}}
\newcommand{\ai}{\mbox{$\alpha _{1}$}}
\newcommand{\aib}{\mbox{$\bar{\alpha _{1}}$}}
\newcommand{\e}{\mbox{$e^{ik_{0}Y}$}}
\newcommand{\at}{\mbox{$\alpha _{2}$}}
\newcommand{\atb}{\mbox{$\bar{\alpha _{2}}$}}
\newcommand{\ath}{\mbox{$\alpha _{3}$}}

\documentstyle[12pt]{article}
\newcommand{\kim}{\mbox {$ k_{1}^{\mu}$}}
\newcommand{\kom}{\mbox {$ k_{0}^{\mu}$}}
\newcommand{\ki}{\mbox {$ k_{1}$}}
\newcommand{\kt}{\mbox {$ k_{2}$}}
\newcommand{\kimb}{\mbox {$ \bar{k_{1}^{\mu}}$}}
\newcommand{\kinb}{\mbox {$ \bar{k_{1}^{\nu}}$}}

\newcommand{\kib}{\mbox {$ \bar{k_{1}}$}}
\newcommand{\ktb}{\mbox {$ \bar{k_{2}}$}}
\newcommand{\qt}{\mbox {$ q_{2}$}}
\newcommand{\qi}{\mbox {$ q_{1}$}}
\newcommand{\qtb}{\mbox {$ \bar{q_{2}}$}}
\newcommand{\qib}{\mbox {$ \bar{q_{1}}$}}
\newcommand{\qo}{\mbox {$ q_{0}$}}
\newcommand{\ha}{\mbox {$\frac{1}{2}$}}
\newcommand{\ko}{\mbox {$ k_{0}$}}
\newcommand{\yim}{\mbox {$ Y_{1}^{\mu}$}}
\newcommand{\yin}{\mbox {$ Y_{1}^{\nu}$}}
\newcommand{\kin}{\mbox {$ k_{1}^{\nu}$}}
\newcommand{\kon}{\mbox {$ k_{0}^{\nu}$}}
\newcommand{\ktm}{\mbox {$ k_{2}^{\mu}$}}
\newcommand{\ytm}{\mbox {$ Y_{2}^{\mu}$}}
\newcommand{\li}{\mbox {$    \lambda_{1}$}}
\newcommand{\lt}{\mbox {$    \lambda_{2}$}}
\newcommand{\lp }{\mbox {$e^{ia\int _{c} k(t) \partial _{z} X(z+at) dt
+ ik_{0} X(z)}$}}
\newcommand{\lpp}{\mbox {$e^{i\int _{c} \alpha (t)
k(t) \partial _{z} X(z+t) dt +ik_{0}X}$}}

\newcommand{\la}{\mbox{$ \lambda $}}
\newcommand{\be}{\begin{equation}}
\newcommand{\br}{\begin{eqnarray}}
\newcommand{\ee}{\end{equation}}
\newcommand{\er}{\end{eqnarray}}

\newcommand{\gvk}{\mbox {$ e^{i\sum _{n > 0}k_{n}Y_{n}}$}}

\newcommand{\eln}{\mbox {$ e^{\sum _{n}\lambda _{n}L_{-n}}$}}

\newcommand{\dsi}{\mbox {$\frac{\partial \sigma}{\partial x_{1}}$}}
\newcommand{\dsib}{\mbox {$\frac{\partial \sigma}{\partial \xb _{1}}$}}
\newcommand{\dsn}{\mbox {$\frac{\partial \sigma}{\partial x_{n}}$}}
\newcommand{\dsq}{\mbox {$\frac{\partial \sigma}{\partial x_{n+m}}$}}
\newcommand{\dst}{\mbox {$\frac{\partial \sigma}{\partial x_{2}}$}}
\newcommand{\dstb}{\mbox {$\frac{\partial \sigma}{\partial \xb _{2}}$}}
\newcommand{\dsth}{\mbox {$\frac{\partial \sigma}{\partial x_{3}}$}}
\newcommand{\dds}{\mbox {$\frac{\delta}{\delta \sigma}$}}
\newcommand{\dsit}
{\mbox {$\frac{\partial ^{2}\sigma}{
\partial x_{1}\partial x_{2}}$}}
\newcommand{\dsb}
{\mbox {$\frac{\partial ^{2}\sigma}{
\partial x_{1}\partial \xb _{1}}$}}
\newcommand{\dsnm}
{\mbox {$\frac{\partial ^{2}\sigma}{
\partial x_{n}\partial x_{m}}$}}
\newcommand{\dsii}{\mbox {$\frac{\partial ^{2}\sigma}
{\partial x_{1}^{2}}$}}
\newcommand{\dsiib}{\mbox {$\frac{\partial ^{2}\sigma}
{\partial \xb _{1}^{2}}$}}
\newcommand{\p}{\mbox {$ \partial$}}
\newcommand{\bj}{\mbox {$ \bar{\partial ^{2}}$}}
\newcommand{\bjj}{\mbox {$ \bar{\partial ^{3}}$}}
\newcommand{\pp}{\mbox {$ \partial ^{2}$}}
\newcommand{\ppp}{\mbox {$ \partial ^{3}$}}

\begin{document}
\title{Some Issues in the Loop Variable Approach to Open Strings
and an Extension to Closed Strings}
\author{B. Sathiapalan\\ {\em Institute for Theoretical Physics},\\
{\em State University of New York at Stony Brook,NY-11794}
\\ and \\
{\em
Physics Department,
Penn State University,} \thanks{Permanent Address}\\{\em 120
Ridge View Drive, Dunmore, PA 18512}}
\maketitle
\begin{abstract}
Some issues in the loop variable renormalization group approach to gauge
invariant equations for the free fields of the open string are discussed.  It
had been shown in an earlier paper that this leads to a simple form
of the gauge transformation law.  We discuss in some detail some
of the curious features encountered there.  The theory looks a little
like
a massless theory in one higher dimension  that can be dimensionally
reduced to give a massive theory. We discuss the origin of some constraints
that are needed for gauge invariance
and also for reducing the set of fields to that of standard
string theory. The mechanism of gauge invariance and the connection
with the Virasoro algebra is a little different from the usual story
and is discussed.  It is also shown  that these results can be
extended in a straightforward manner to closed strings.
\end{abstract}
\newpage
\section{Introduction}
  One of the most important issues in string theory that needs elucidation
is that of space time symmetries of the theory.
Closed string theory, in particular,
contains general coordinate invariance in its low
energy limit and we do not know what the appropriate string
generalization of this is. The BRST formalism\cite{SZ,EW,Z}
that has been used succesfully to
construct a non polynomial closed string field action probably
contains the answer to this question.  But at the moment it remains
hidden deep inside.  A formalism where the answer to this question
is manifest is a crying need at this point - if only because we are
unlikely to be able to do any physically significant computation
otherwise. In particular the problem of background dependence
of most formalisms becomes an impediment to non perturbative calculations
{}.

In \cite{BSI} we had speculated that some generalization of the
renormalization group (in space-time) could be the space time
interpretation of the symmetries of string theory.  Motivated by
this idea, we were able to describe the open string gauge
transformation as a local ( in $\sigma$ ) scale transformation.
Furthemore we were able to derive the gauge invariant equations of the
of the fields of the open string using a generalization of the
sigma model $\beta $ function ( 2-dimensional renormalization group)
methods developed earlier \cite{LFT,CSen,DS,BSII}.  This generalization
involves using a nonlocal loop variable defined
on the boundary of a circular hole in the world sheet rather than the
usual vertex operaror located at a puncture.  As argued in \cite{BSII,BM}
while going
off shell in the sigma model approach one should maintain a finite
cutoff.  If one keeps a finite cutoff one must include all the
irrelevant operators which, in string theory, correspond to the vertex
operators of the massive modes. Furthermore one would
like to do all this in a gauge invariant way.  We believe that the
loop variable approach \cite{BSI} shows some promise in this regard.

  However there are many curious
features in that formalism that require further analysis.  One is the
contrast between the form of the gauge transformation used in
\cite{BSI} and that generated by the Virasoro generators (or $Q_{BRST}$)
in usual formulations of string theory.  Some aspects are
discussed in \cite{BSIII}. A related puzzle was the fact that the gauge
transformations in \cite{BSI} were of the standard form that one gets in
Kaluza Klein theories: Start with a massless, D+1 dimensionsal,
gauge invariant theory and dimensionally reduce to get a gauge invariant
massive D-dimensional theory.  In the usual formulations\cite{SZ,EW}, in
order to get the action and transformations to the standard form
one has to perform field redefinitions at each mass level.
Furthermore they work only in 26 dimensions (for the bossonic string).
The method of \cite{BSI}  does not require any field redefinitions.
Since we are requiring conformal invariance we are restricted to D=26.
However this does not show up in a direct way as in \cite{SZ}
Somehow it is as if the
field redefinitions have been implicitly performed at the beginning
itself.  Of course as an algorithm for deriving gauge invariant equations for
massive fields it works in any dimension.  But only in
D=26 can one make contact with standard string theory. In other
dimensions it is not the standard string theory.
The connection with the usual formulations of string theory
and the question of field redefinitions, we believe, is the
most pressing one although we will not have much to say about this
in the present work.

 A related feature is the partially symmetric treatment of the
extra ghost coordinate.  Counting arguments \cite{SZ}have shown that all
the extra degrees of freedom necessary for the string field
can be obtained by adding one extra bosonic oscillator
minus its zero mode and first mode.  In \cite{BSI} we added an extra
 coordinate , but integrated it out and replaced it with the
Liouville mode.  This was motivated by the work of \cite{CSen}.
It turns out that this is unnecessary and one can treat it just
as an extra dimension.  Thus we seem to have a D+1 dimensional theory.
However in making contact with the vertex operators of string
theory it becomes necessary to impose some constraints on the
generalized momenta.  It turns out that these constraints
are essentially the ones we were forced to impose by hand
in \cite{BSI} in order to have gauge invariance.  These constraints
were, in fact, another of the curious features of that method.

 The gauge transformations in \cite{BSI}
had the form of scale transformation
.  There was no obvious restriction that it had to be infinitesimal.
Yet it was the lowest order terms that we kept and which were
equivalent to the usual gauge transformations of string theory.
The obvious question is whether the next order terms give
anything new.  It turns out that the answer is in the negative.  However
in the process of understanding this we will also
understand how computationally the gauge invariance actually works
and also the origin of a tracelessness condition on the gauge parameters.

Finally,  it is not immediately obvious
whether these techniques work equally well for the closed string.
We will show that at least for the lowest mass level they do work.

In this paper we address some of these issues.  In Sec.II we
briefly review the results of \cite{BSI} and describe how one can
simplify many features by treating the extra coordinates in a symmetric
fashion and by not integrating it out as was done there.
In Sec. III we discuss the
connection between the gauge transformation there and the action
of the Virasoro generators as discussed in \cite{BSIII,BSIV}.
In Sec. IV we discuss the mechanism of gauge invariance and also discuss
 the issue of higher order terms.  In Sec. V we extend
the discussion to closed strings.

\newpage
\section{Loop Variables}
In this section we briefly review the results of \cite{BSI} and then give a
modified simpler treatment.   The basic idea there is to consider
a generalization of the vertex operator in the form of a loop variable.
\be
\lp
\ee
The integral is along a contour 'C' which
is taken to be a circle of
radius '$a$' centered at the point $z$.
As explained in \cite{BSIV} this loop
variable is actually more general than just the collection of vertex
operators that is obtained when one Taylor expands $X(z+at)$
in positive powers of '$at$'.  However we will ignore this for the
moment and Taylor expand $ \partial X(z+at)$ and $k(t)$ in powers
of $t$ as follows:
\br
\p X(z+at) & = & \p X(z) + at \pp X + (at)^{2} /2 \ppp X + ...
 (at)^{n}/n! \p ^{n+1} X +... \nonumber \\
k(t)       & = & k_{0} + k_{1}/t + k_{2}/t^{2} + ... +k_{n}/t^{n} +...
\er
Just as the vertex operator is a puncture on the world sheet, the loop
variable is the boundary of a circular hole of radius '$a$' in the
world sheet, centered at the point '$z$'.  A point on the loop is
$z'= z+ at = z+ ae^{i\theta }$.
One has to allow for reparametrizations of $z'$ that leave the shape
of the boundary unchanged,i.e., maps that map $z'$ at one point on the
boundary to another point on the boundary \cite{AMP,OA,CMNP,R}.
In order to implement this we introduce an einbein $\alpha (t)$ so
that equation (2.1) is modified to : (We have set $a=1$)
\be
\lpp
\ee
and we assume that $\alpha (t) $ has an expansion
\be
\al (t) = \al _{0} (=1) + \ai /t + \at /t^{2} +... + \al _{n} /t^{n} +..
\ee
Substituting the power series expansions (2.2) and (2.4) into (2.3) we get
\be
e^{i(k_{0}Y + k_{1} Y_{1} + k_{2}Y_{2} + k_{3 }Y_{3} +...)}
\ee
where $Y,Y_{1},Y_{2}...Y_{n}$ are defined as follows:
\be
Y= X+ \ai \p X + \at \pp X + \ath \ppp X /2! +...,
\ee
Define the variables $x_{m}$ by the relation :$\frac{\partial
 \alpha _{n}}{\partial x_{m}} = \alpha _{n-m} ; \alpha _{0} =1$.
Then
\be
Y_{n} = \frac{\p Y}{\p x_{n}} = \p ^{n}X /(n-1)! + \ai \p ^{n+1}X/n! +...
\ee
The expression (2.5) is the basic object we work with.  Vertex operators
are obtained by expanding the exponential in (2.5).  Thus $e^{ik_{0}Y}$
is the tachyon operator, $ik_{1}Y_{1}e^{ik_{0}Y}$ is the spin one
vertex operator, $ik_{2}^{\mu}Y_{2}^{\mu} e^{ik_{0}Y}$ and $
-\frac{1}{2}k_{1}^{\mu}k_{1}^{\nu} Y_{1}^{\mu}Y_{1}^{\nu} \e$
are the vertex operators at the next mass level.  The vertex operators
in (2.5) contain an anomalous dependence on the Liouville mode.  We
can obtain a compact expression for this by defining a new Liouville
field:
\be
2 \sigma _{new} = <YY> = <XX> +2 \ai <\p X X> +...\equiv  2 \sigma +
\ai \p \sigma +...
\ee
Then using
\br
<Y_{n} Y> & = & 1/2 \frac{\p}{\p x_{n}} <YY> =  \dsn \nonumber \\
<Y_{n}Y_{m}>& = & 1/2(\dsnm - \dsq )
\er
we get on including the Liouville mode dependence, in (2.5):
\be
e^{ik_{n}Y{_n} - \frac{1}{2}k_{n}.k_{m} (\dsnm - \dsq )}
\ee
Finally the fields are obtained as follows:
\br
\phi (x) & = & \int [dk_{n}] \Phi [k_{n}] \e \nonumber \\
\phi _{1} ^{\mu} (x) & = & \int [dk_{n}] \kim \Phi [k_{n}] \e  \nonumber
\\
\phi _{11}^{\mu \nu} (x) & = & \int [dk_{n}] \kim \kin \Phi [k_{n}] \e
\er
and so on.  Here $\Phi [k_{n}]$ is the string field.
The gauge invariant equations are obtained by requiring
\be
\dds \int[dk_{n}]\Phi [k_{n}]
e^{ik_{n}Y_{n} - \frac{1}{2}k_{n}.k_{m} (\dsnm - \dsq )}  =0
\ee
In order to get the right $(mass)^{2}$ one must remember to replace the
radius $'a'$ of the loop by $ae^{2\sigma}$.
So far we have not discussed the ghost coordinate.  In \cite{BSI} we
had included an extra space time coordinate $\theta$ and a conjugate
momentum $q$, so the vertex operator (2.5) becomes
\be
e^{i k_{n}^{\mu} Y_{n}^{\mu} + iq_{n} \theta _{n} }
\ee
We then integrated out $\theta _{n}$ and were left with the
induced Liouville terms as in (2.10).  We used $\frac{\partial ^{2}
\sigma }{\partial x_{n} \partial x_{m}} \e$
as the vertex operator corresponding to the auxiliary degrees of
freedom.  Thus we treated the extra coordinate somewhat asymmetrically
, but not as asymmetrically as the bosonized ghost coordinate should
be treated - since the action for the ghost coordinate contains
a coupling to the world sheet curvature scalar - which we do not have.
The precise connection between $\theta$ and the bosonized ghost is
something that needs to be worked out.  In any case we will see below
that the treatment of $\theta$ can,in fact, be completely symmetric.
Thus we can just use the vertex operator (2.10) and require (2.12) to hold
without any consideration of extra auxiliary coordinates.  The equations
have the invariance
\br
k(t) \rightarrow k(t) \la (t)   \nonumber \\
\la (t) = \la _{0} + \li /t    + \la _{2} /t^{2} +...
\er
just as in \cite{BSI}.  These equations are precisely those for massless
D+1 dimensional gauge fields of different spin! At this
point the connection
with string theory is completely obscure since this is
hardly the right spectrum.
However one can 'dimensionally
reduce' a la Kaluza-Klein and set the D+1 st component of momentum
equal to the mass to get the
right spectrum.
However we must remember that in string theory
it is the $(mass)^{2}$ that is an integer multiple and not the mass,
as would be obtained if we really set the internal
momentum equal to a mass.  We will just set the variable
$\qo = mass $ by hand to get the right spectrum.  We are
effectively setting $\qo ^{2}$ equal to the naive
dimension of the vertex operator.  This is equivalent
to replacing the radius '$a$' by $ae^{2 \sigma }$ in eqn.(2.2),
which is just what is required on a curved world sheet \cite{AMP,BSI}.
However we lose the conventional geometric interpretation
of $\theta$ as a D+1 st compactified space coordinate.
We will illustrate this now for the level two fields
.  The vertex operators at the second mass level are:
\be
e^{\ko ^{2} \sigma  + i \ko Y}[-1/2! \kim \kin \yim \yin  + i \ktm
 \ytm + i \ki \ko \dsi \kim \yim +
\ee
\[
\ha \ki . \ki ( \dsii - \dst )
+ \kt . \ko \dst ]
\]
Setting $\frac{\delta}{\delta \sigma}$[expression (2.15)]$= 0$ we get
the following two equations:
\be
[-1/2 \ko ^{2} \kim \kin + \ki . \ko \kim \kon - 1/2 \ki . \ki
\kom \kon ] \yim \yin \e = 0
\ee
\be
[i\ko ^{2} \kim \kin + \ki . \ko \kim \kon - 1/2 \ki . \ki \kom
-i \kt . \ko \kom ] \ytm \e = 0
\ee
These equations are invariant under the gauge transformations (2.14):
\be
\ki \rightarrow \ki + \li \ko  \, ; \, \kt \rightarrow \kt + \li \ki
+ \la _{2} \ko
\ee
Let us set $k^{D+1}(t)=q(t) , \, Y^{D+1} = \theta $ and let the internal
momentum  $q_{0}$ be the mass.  Then we get gauge invariant equations
for a set of massive 'spin'2 and 'spin' 1 fields.  Equation (2.16)
 reduces to a set of three equations:
\[
[-1/2(\ko ^{2} + \qo ^{2} ) \kim \kin + ( \ki . \ko + \qi \qo )
1/2( \kim \kon + \kin \kom )
\]
\[
- 1/2 ( \ki . \ki + \qi \qi ) \kom \kon ]=0
............(2.16a)
\]

\[
 [-1/2 (\ko ^{2} + \qo ^{2} ) \kim \qi + 1/2 (\ki . \ko + \qi \qo  )
( \kim \qo + \qi \kom )
\]
\[
-1/2 (\ki . \ki + \qi \qi ) \kom \qo ] =0
....... .........(2.16b)
\]

\[
[-1/2 ( \ko ^{2} + \qo ^{2} ) \qi \qi +
(\ki . \ko + \qi \qo ) \qi \qo
- 1/2 ( \ki . \ki + \qi \qi ) \qo ^{2} ]
\]
\[
= [ -1/2 \ko ^{2} \qi \qi + \ki \ko \qi \qo - 1/2 \ki . \ki \qo ^{2} ]
=0.............(2.16c)
\]
which are the coefficients of $Y_{1}^{\mu}Y_{1}^{\nu} ,\, Y_{1}^{\mu}
\theta _{1} , \, and \, \theta _{1} \theta _{1} $ respectively.
Similarly (2.17) reduces to:
\[
[(\ko ^{2} + \qo ^{2} ) \ktm - ( \ki .  \ko + \qi \qo ) \kim
 + ( \ki . \ki + \qi \qi ) \kom - ( \kt . \ko + \qt \qo ) \kom ] =0
.......... (2.17a)
\]

\[
[( \ko ^{2} + \qo ^{2} ) \qt - ( \ki . \ko + \qi \qo ) \qi +
( \ki . \ki + \qi \qi ) \qo - ( \kt . \ko + \qt \qo ) \qo ] =
\]
\[
[\ko ^{2} \qt - \ki . \ko \qi + \ki . \ki \qo - \kt . \ko \qo ]
=0.............(2.17b)
\]
which are the coefficients of $Y_{2}^{\mu} $  and $ \theta _{2}$
respectively.

Equations (2.16a-c) describe a massive spin 2 field with
the requisite auxi
 liary fields ( a vector and a scalar ).  Thus $\phi _{11} ^{\mu , \nu }
\sim k_{1}^{\mu} k_{1}^{\nu} $ ;  $\phi _{11}^{\mu 5} \sim k_{1}^{\mu}
q_{1}$ ; $\phi _{11}^{55} \sim q_{1}q_{1}$ describe these fields.
The gauge transformations (using 2.18) are:
\be
\delta \phi _{11} ^{\mu \nu} = \epsilon _{2} ^{(\mu} \ko ^{\nu )}
\, ; \, \delta \phi _{11} ^{\mu 5}= \kom \epsilon _{2} + \epsilon ^{\mu}
\, ; \, \delta \phi_{11}^{55} = 2 \epsilon _{2}
\ee
where we have set $\lambda _{1} k_{1} ^{\mu} \sim \epsilon _{2}^{\mu}$,
$\lambda _{1} q_{1} \sim \epsilon $ \footnote{See Sec.III
for more details}.
  Thus $\phi _{11}^{\mu 5}$ and $ \phi _{11} ^{55}$ can be gauged
away and one recovers the Fierz Pauli equation for a massive spin 2
field. It is important to note that no field redefinitions are
required.

Equation (2.17a) describes a massive spin one field \ktm in addition
to \kim $q_{1}$.
In fact this was the equation we had in \cite{BSI} where we made the
identification $\kim q_{1} \sim \ktm$.  With this identification
(2.17a) and (2.16b) are in fact the same equation.  This condition
amounts to saying that $'q_{1}'$ is not an entirely independent
degree of freedom.  In \cite{SZ} it was shown by a counting argument
that one should set the zeroeth and first modes of a bosonic field
to zero in order to reproduce the auxiliary field counting.  What
we are seeing here is just another version of that.  In fact (2.16c)
becomes identical with (2.17b) if we further set $q_{1}q_{1}
\sim q_{2}$.  Furthermore on adding the trace of (2.16a) with (2.16c)
one reproduces the equation associated with the vertex operator
$\frac{\partial ^{2} \sigma }{\partial x_{1}^{2}}$ in \cite{BSI}.

The vertex operator $\ktm \partial ^{2} X^{\mu}$ is not covariant
when one has a curved background - in fact one should have
$(\partial ^{2} X + \partial \sigma \partial X)$.  If we think of
$\theta  $ as somehow representing the degree of freedom $\sigma $
then the vertex operator $(\ytm + \theta _{1} Y_{1}^{\mu})$ is the
vertex operator at the second mass level.
 Thus it makes sense to identify $ \kim q_{1} $ with $\ktm $.
We must conclude that the identification $\kim q_{1} \sim \ktm$
is required if one is to make contact with the usual vertex operators
 of string theory.  If we do not make these identifications we seem to
have a simpler theory, but with a bigger set of vertex operators
 and fields. In that case the counting of states would be quite different
from that of the usual BRST formalism.

To summarize this section, if we treat the extra 'ghost' coordinate
in string theory as an extra dimension and write the theory as a
D+1 dimensional theory we get a
set of equations for a massless spin 2 field with an auxiliary field.
If we now dimensionally reduce to D dimensions, the three vertex
operators $\yim \yin $, $\yim \theta _{1}$
and $(\theta _{1})^{2}$ give us all
the fields necessary for a covariant massive spin 2.  This is similar
to the
usual Kaluza Klein phenomenon.  The vertex operators \ytm and
$\theta _{2}$ then give an extra set of vector and scalar fields.
The theory is still gauge invariant.  If we now impose $\kim q_{1}
\sim \ktm$ and $q_{1}q_{1} \sim q_{2}$
we truncate back to the usual set
(spin 2, spin 1, spin 0) for a gauge invariant, manifestly Lorentz
invariant massive spin 2 and get rid of the extra fields.
We have the same counting of states as in BRST formalism of
string theory and
the same equations of motion.
However it would be nice to have a formal proof
that this construction is equivalent to all orders,
in 26 dimensions, (upto  field redefinitions) with string theory.

\newpage
\section{Gauge Transformations}
\setcounter{equation}{0}

   In this section we discuss the gauge transformations in \cite{BSI}
 and the
connection with Virasoro generators.  In \cite{BSI} we had the string field
(we suppress the einbein $\alpha (t)$ for the moment).
\be
\int {\cal D} k(t) \lp \Phi [k(t)] = \Phi [X(C)]
\ee
The usual point particle fields are obtained from (3.1) as in (2.11).
$ \Phi [X(C)]$ thus is not the usual string field which ,in fact, is
obtained by acting with $\Phi$ on the (SL(2,R) invariant) vacuum.
Thus
\be
\Psi [X(C)] = \Phi [X(C)] \mid 0>
\ee
is the object commonly referred to as the string field.
The transformation properties of $\Phi$ under
the action of Virasoro generators are
clearly different from those of $\Psi$ since the vacuum is only invariant
under $L_{-n}$ with $n=0,\pm 1$\footnote{In the BRST formalism $Q_{BRST} \mid
0>
=0$.  It would seem that this distinction is irrelevant.  Nevertheless
the gauge transformation in the BRST formalism has terms corresponding, in the
covariant formalism, to terms of the form $L_{-n} \mid 0>$.  See below.}
.  These were worked out in \cite{BSIII,BSIV}
in some detail.  It was shown there that the action of $L_{-n}$,
$n>0$ on the loop variable (2.5) has two parts to it:  The first piece
is obtained by making the transformation:
\br
k_{m+n} & \stackrel{L_{-n}}{\rightarrow }  & k_{m+n} +
m \la _{n} k_{m} \nonumber \\
k_{n}   & \stackrel{L_{-n}}{\rightarrow }  & k_{n}  + \la _{n} k_{0}
\nonumber \\
k_{m}   & \stackrel{L_{-n}}{\rightarrow }  & k_{m}  \, , n>m
\er
This amounts to saying that $k(t)$ transforms as a scalar and $\partial
_{z} X(z+t)$ as a vector (or a tensor of weight one) under the
diffeomorphism
\be
t\rightarrow t(1+ \lambda (t))
\ee
which is clearly different from (2.14).  On the other hand let us start
with the loop variable
\be
e^{i \int _{c} \frac{dt}{t} k(t) X(z+t) }
\ee
This can be obtained from (2.1),(3.1) by rescaling the $k_{n}$'s:

\be
k_{n} \rightarrow k_{n}/n
\ee
Thus the transformations (3.3) becomes
\be
k_{m+n}  \stackrel{L_{-n}}{\rightarrow }   k_{m+n} + (m+n) \la _{n}
 k_{m}
\ee
which is equivalent to $k(t)/t \rightarrow k(t)/t + \frac{d}{dt}
[t\lambda (t) k(t)/t]$.  (3.7) can be obtained by transforming '$t$'
in $X(z+t) $  according to
\be
t \rightarrow t(1+ \la (t) ) = t + \la _{0} t + \li + \lt /t +...
\ee
and requiring invariance of (3.5).
Since $t\la (t)$  is the change in $t$, this is equivalent to
requiring that $k(t)/t$ transform as a tensor of weight one.  Thus
\be
e^{i \int _{c} dt/t k(t) X(z+t) }  \stackrel{\delta (t)}{\rightarrow}
e^{i \int _{c} dt/t k(t) X(z+t + \delta (t) ) }
\ee
\[
= e^{i \int _{c} dt/t k(t) X(z+t) }
e^{i \int _{c} dt \delta (t) /t k(t) \p X(z+t) }
\]
for infinitesimal $\delta$.  Thus when $\delta (t) /t =1$, the
second factor is the loop variable (2.1).  If we now change $\delta (t)/t
\rightarrow \delta (t) \la (t) /t $, then the second factor becomes
(for $\delta (t)/t =1$), the gauge transformed loop variable.  Thus
the $k(t) \rightarrow \la (t) k(t) $ can be viewed as the transformation
induced on (2.1) by the action of a Virasoro generator not on (2.1) but
on the {\em loop variable (3.5)}.  Heuristically, if we think of the loop
variable (3.5) as a group element  $g[k]$, then the Virasoro generators
induce  a transformation
\be
g[k] \rightarrow g[k+ \delta k ]  \equiv g[k]h[\delta k]
\ee
The loop variable (3.1) is obtained by translating back to the origin by
$g^{-1}$:
\be
g^{-1}[k]g[k+ \delta k] = h[ \delta k]
\ee
[To lowest order in $\delta$ , $h[\delta k] = 1+ g^{-1} \delta g$].
The gauge transformation (2.14) corresponds to changing $\delta k
\rightarrow \la \delta k$ so that
\be
h[\delta k] \rightarrow h[ \la \delta k]
\ee
Thus the transformation (2.14) while not a diffeomorphism of $S^{1}$,
is induced by one in the manner outlined.  It has a simpler space time
interpretation, however - it is a rescaling of $k(t)$ and hence of $X(t)$
.  Since it is a pointwise multiplication, the operation $k(t)
\rightarrow k(t) \la (t) $ has a simple geometrical interpretation (of
local rescaling) and the parameter along the string,$t$, has no role
to play- which is just as well since it has no physical significance.
This is not the case in diffeomorphisms - where the transformation
involves the derivatives with respect to $t$.

So far we have discussed only one part of the action of the Virasoro
group on vertex operators \cite{BSIII,BSIV}.  When one calculates the operator
product (OP) of the group element \eln on a vertex operator
\gvk , we are , in fact, computing the action of \eln on the
state created by \gvk,i.e.
\be
\eln \gvk \mid 0>
\ee
and from \cite{BSIV} this has the form (to lowest order in \la )
\be
e^{1/2 Y_{n} \lambda _{n+m} Y_{m} -1/2 nk_{n} \lambda _{-n-m} mk_{m}
-ink_{n} \lambda _{-n+m}Y_{m}}
\ee
The first term in the exponent comes from the action of the $L_{-n}$
on the vacuum state $\mid 0>$ since it is not invariant.  The
second term comes from the $L_{+n}$ and these are constraints
rather than gauge transformations and will not concern us here.  The
third term is the action of $L_{-n}$ on the vertex operator.  From
our discussion thus far it is clearly the last term that is involved
in the gauge transformation of the "string field" in \cite{BSI}.
That is we have been defining the transformation of the space time
fields by transforming $\Phi [X(C)]$ rather than $\Psi [X(C)]
= \Phi [X(C)] \mid 0>$.  In the usual formalism it is the
transformation of the latter field that defines the gauge transformation
on the space time fields.  In fact it was noted in \cite{BSI} that the gauge
transformation in the BRST formalism \cite{SZ} had an extra $\delta ^{\mu \nu}
$ piece which was then gotten rid of by means of field redefinitions.
This extra term is precisely of the form of the first term in the
exponent of (3.14).  This is one important difference between the formalism
of \cite{BSI} and the usual string field theory formalism.  The precise
mechanism by which, in 26 dimensions, one can perform field redefinitions
that take one from $\Psi $ to $\Phi$ needs further elucidation.  Some
suggestions were made in \cite{BSIII}.

What we have discussed thus far in this section is the connection
between gauge transformations as defined by equation (2.14) and
diffeomorphisms of $S^{1}$.  Now starting from (2.14), with the
vector field defined as in equation (2.11) to get to the gauge
transformation law:
\be
\phi _{1} ^{\mu} ( \ko ) \rightarrow
\phi _{1} ^{\mu} ( \ko ) + \kom \Lambda
\ee
we have to assume that the string field
$\Phi [k_{n}]$ depends on \la     as
well as $k$ so that (3.15) comes from
\be
\int[dk_{n}][d \la ] \Phi [k, \la ] \kim \rightarrow
\int[dk_{n}][d \la ] \Phi [k, \la ] ( \kim + \li \kom )
\ee
with
\be
\int [dk][d \la ] \Phi [ k, \la ] \li \equiv \Lambda
\ee
the gauge parameter.
The inclusion of a \la  -dependence inside $\Phi$ is unusual and has
no counterpart in usual field theory, where the gauge parameter is
separate from the fields.  In our case $ \la (t)$ multiplies
$\alpha (t)$ the einbein.  Thus one could just as well think of the
einbein as parametrising the space of gauge transformations and then
it is not unnatural to let the string field depend on $\alpha (t)$,
which could play the role of the ghosts.
However this issue of the $\la $ dependence of the string field
needs further investigation.

\newpage
\section{Mechanism of Gauge Invariance}
\setcounter{equation}{0}

   In this section we describe the mechanism of gauge invariance and
also discuss what happens at higher orders in the gauge parameter
\la .  As described in \cite{BSI} we have to start with the loop variable
(2.3), with an einbein $\alpha (t)$, which is integrated over since
one has to sum over reparametrizations of the boundary.  Thus we have
\be
\int {\cal D} \alpha (t) \lpp
\ee
Clearly at the formal level this is invariant under $\alpha (t)
\rightarrow \alpha (t) \la (t) $ since this is just a change of
variables, provided the measure ${\cal D} \alpha (t) $
is invariant. In \cite{BSI} we set \footnote{It is conceivable that this is
true
only in 26 dimensions}
\be
{\cal D} \al (t) = [ dx_{1} dx_{2} ...] \, ; \, \al (t) =
e^{\sum _{n} x_{n} t^{-n} }
\ee
Thus if we parametrise  $\la (t)$ by:
\be
\la (t) = e^{ \sum _{n} y_{n} t^{-n}}
\ee
clearly a gauge transformation corresponds to a translation of the
$x_{n}$'s:
\be
x_{n} \rightarrow x_{n} + y_{n}
\ee
under which the measure (4.2) is invariant.  However while this is formal
ly true (4.1) contains an implicit dependence on $\sigma $, reflecting
the ultraviolet divergences that come about when doing the X-functional
integral.  Let us write all the terms upto $Y _{3}$:
\be
exp(i(\ko Y + \ki Y_{1} + \kt Y_{2} + k_{3} Y_{3} ))
exp(\ko ^{2}\sigma + \ki . \ko \dsi + \kt . \ko \dst +
\ee
\[
\ki . \ki /2
(\dsii - \dst ) + k_{3} . \ko \dsth + \kt . \ki ( \dsit - \dsth ))
\]
We can rearrange the terms as follows:
\be
exp(i[\ko . (\ko \s + iY) +
\ki \frac{d}{dx_{1}} ( \ko \s + iY) +
\kt \frac{d}{dx_{2}} ( \ko \s + iY)  +
\ee
\[
k_{3} \frac{d}{dx_{3}} ( \ko \s + iY) + \ki . \ki /2 ( \dsii - \dst )
+ \kt . \ki ( \dsit - \dsth )])
\]
Furthermore using
\be
\frac{d^{3}}{dx_{1}^{3}} \s  =
\frac{d^{3}}{dx_{1}^{3}} <YY>  = 2[<Y_{2} Y> +3 <Y_{2} Y_{1}>]=
-2 \dsth +3 \dsit
\ee
we see that
\be
\frac{d}{dx_{1}} ( \dsii - \dst ) =
2( \dsit - \dsth )
\ee
Thus we can further rewrite (4.6):
\be
exp(i[( \ko +
\ki \frac{d}{dx_{1}}  +
\kt \frac{d}{dx_{2}}  +
k_{3}\frac{d}{dx_{3}} )(\ko \s + iY)
\ee
\[
+ \ki . \ki /2 ( \dsii - \dst )
+ \ki . \kt \frac{d}{dx_{1}} ( \dsii - \dst )])
\equiv e^{A}
\]
If we make gauge transformations with parameter $\la _{1}$ we get:
\be
exp(A + i \li \frac{d}{dx_{1}} (\ko Y + \ki Y_{1} + \kt Y_{2} )
\ee
\[
+\li \frac{d}{dx_{1}}(\ko ^{2} \s + \ki .\ko \dsi + \kt . \ko \dst
 +\ki . \ki ( \dsii - \dst ) )
\]
\[
 = e^{A + \li \frac{dA}{dx_{1}}}
\]
to this order.
{}From (4.3) we see that to lowest order $\la _{1} = y_{1}$ and thus we
see that all we have to this order is a transformation
of $x_{1}$ by an
amount $\la _{1} = y_{1}$ - which is just what we expect from (4.4).
When we derive the equations of motion by varying respect to $\sigma $
we routinely integrate by parts on the $x_{n}$.  This of course
corresponds to adding total derivatives -
which we are allowed to do inside
 an integral (assuming no boundary terms).  Thus we rewrite
$\la _{1} \frac{d}{dx_{1}}A $ as $\sigma B$,for some $B$.  Now
$\la _{1} \frac{d}{dx_{1}}A $ is a total derivative in $x_{1}$ and so
must $\sigma B$ be one [since integrating by parts corresponds
to adding more total derivatives].  Clearly $\sigma B$ cannot be a total
derivative since there is no derivative on $\sigma$.  Thus the only
possibility is B=0.  Thus we see that a $\sigma $ variation of a gauge
 variation of A is zero.  Since a $\sigma $ variation commutes with a
gauge variation we would seem to have proved that the gauge variation
of an equation of motion [i.e. a $\sigma $ variation  of A] is zero.
  While this argument is true as it stands one has to be a little
careful about applying it here since it involves using (4.8).  Identities
like (4.8) and (4.7) can be used to rexpress $\frac{\delta}{\delta \sigma
} A$ in terms of different sets of vertex operators.  For instance
$\frac{\delta}{\delta \sigma } \frac{\partial ^{3} \sigma}{\partial
x_{1}^{3}}\e$ will have a term involving $\yim \yin Y_{1}^{\rho}$
whereas $\frac{\delta}{\delta \sigma}(-2\frac{\partial \sigma}{\partial
x_{3}} + 3\frac{\partial ^{2} \sigma}{\partial x_{1} \partial x_{2}})
\e$ will have only terms quadratic in derivatives of $Y$.  Thus the
argument made earlier can be applied
to the {\em sum} of all terms in the
variation of A (i.e. including different vertex operators). However
if we want to set the coefficient of each vertex operator separately
to zero our gauge invariance is necessarily more restricted.  Thus the
gauge variation of the terms of the type in (4.8) must necessarily be
zero.  Now $\delta (k_{1}k_{2}) = \la _{1} k_{1}.k_{1} + \la _{1}
k_{2}.k_{0}$.  It is the former term that involves the vertex operator
$\frac{\partial ^{2} \sigma }{\partial x_{1}^{2}}$, as can be seen
in (4.10).  Thus we must require
\be
\int [d\la dk] \li \ki .\ki \Phi [k, \la ] =0
\ee

This amounts to saying  that the gauge parameter is traceless.  This
condition is familiar in higher spin gauge theories.  If we look at the
gauge invariance of the spin 3 equation of motion, i.e. the coefficient
of $\yim \yim Y_{1}^{\rho}$, we will see that this condition is indeed
required.  This equation is:
\be
[-i\ko ^{2}/3! \kim \kin \ki ^{\rho } + i/2! \ki .\ko \kim \kin \ko
^{\rho } -i \ki .\ki /2 \kim \kon \ko ^{\rho} ] \yim \yin Y_{1} ^{
\rho} =0
\ee
and can be seen to be invariant when the tracelessness condition
is imposed.

Having described the mechanism of gauge invariance, we can easily see
that it can be extended to higher orders in \la .

In eqn. (4.4) if we let $y_{1}$ be finite then one has higher order terms
 in eqn.(4.3).  Thus we get
\be
1+ \li /t + \lt /t^{2} + \la _{3} /t^{3} +...= 1+ y_{1}/t +y_{1}^{2}/2!
1/t^{2} + y_{1}^{3}/3! 1/t^{3} +...
\ee
from which we conclude that
\be
y_{1}=\li  \, ; \, \lt = \li ^{2}/2! \, ; \, \la _{3} = \li ^{3}/3!  ...
\ee
and the gauge transformation on the $k_{n}$ become:
\be
\ki \rightarrow \ki + \li \ko \, ; \, \kt \rightarrow \kt + \li \ki +
\li^{2}/2! \ko
\ee
If we start with (4.6) (or 4.5) and make the above substitution keeping
all terms to $O(\la _{1} ^{2})$ we will find, as expected, instead of
(4.10)
\be
exp(A + \li \frac{dA}{dx_{1}} + \li ^{2}/2! \frac{d^{2}A}{dx_{1}^{2}}+.).
\ee
In the spin 2 equations (17) one can make the substitution (4.15) and
check that they are invariant order by order in \la .  Note that the
field $\phi _{11}^{\mu \nu}$ transforms as:
\be
\int [dk d \la ] \kim \kin \Phi [k, \la ] \rightarrow
\int [dk d \la ] ( \kim \kin + \li ( \kim \kon + \kin \kom ) +
\li^{2} \kom \kon ) \Phi [k, \la ]
\ee
\[
\Phi _{11}^{\mu \nu} \rightarrow \Phi _{11}^{\mu \nu } +
\p ^{( \mu } \Lambda _{11}^{\nu )} + \p ^{\mu} \p ^{\nu } \Lambda _{11}
\]
Thus the higher order pieces in $\la _{1}$ do not result in a
higher order term in
$\Lambda _{11}^{\mu}$
but in a modification of
$\Lambda _{11}^{\mu }$  to $\Lambda _{11}^{\mu} + \partial
^{\mu} \Lambda _{11}$.
Clearly (as expected) this does not result in any new symmetry.  One can
also check the spin 3 equations and as before we need the tracelessness
condition $ \li k_{1}.k_{1}=0$(4.11), but now also a condition
$\la _{1}^{2} k_{1}.k_{0}=0$.  This second condition follows by a gauge
transformation of (4.11) so it is not unexpected.

In this section we have described the basic mechanism of gauge
invariance: the freedom to add total derivatives in the '$x_{n}$'.
This generalizes the well known freedom in $\sigma $-models to add
derivatives in $z$.  We have also described the origin of the
tracelessness condition and also what happens at higher orders
in $\la (t)$.  One issue that needs further exploration  is that of the
extra degrees of freedom $\alpha (t) , \, \la (t)$ or $x_{n}, y_{n}$.
  From (4.4) it is obvious that the only justification for treating
$y_{n}$ differently from $x_{n}$ is that $y_{n}$ is a gauge
{\em parameter} describing the change in $x_{n}$.  On the other hand we
are forced to include $y_{n}$ as a degree of freedom ( as explained in
 Sec. III).  Clearly there must be some way to get around this
duplication.
\newpage
\section{Closed String}
\setcounter{equation}{0}

  In this section we extend the results of \cite{BSI} to closed strings.  The
vertex operators for closed strings are obtained by combining
holomorphic and anti holomorphic open string vertex operators of the
same mass level.  All we have to do then is to apply the construction
described in Sec.II and III separately to the holomorphic and
anti holomorphic generalized vertex operators.  Thus we can generalize
the loop variable to
\be
exp(i\int _{c} \alpha (t) k(t) \partial _{z} X(z+t, \zb + \tb ) dt +
\int _{c} \alb   (\tb ) \bar{k}(\tb )\partial _{\zb} X(z+t, \zb + \tb )
d \tb + ik_{0}X)
\ee
We will Taylor expand the exponent in powers of $t, \bar{t} $, keeping
in mind that $\partial _{z} \bar{\partial _{\bar{z}}} X(z,\bar{z})=0$.
Thus the exponent becomes :
\be
\ko [X(z, \zb ) + \ai \p X + \a2 \pp X + \al _{3} \ppp X/2! +...
+ \aib \bar{\partial} X + \atb \bj X + \bar{\al _{3}} \bjj X/2! +...]
\ee
\[
+ \ki [ \p X + \ai \pp X + \at \ppp X /2! +..]
+ \kib [ \bar{\partial} X + \aib \bj X + \atb \bjj X /2! +...]
\]
\[
+ \kt [ \pp X + \ai \ppp X /2! +...]
+ \ktb [ \bj X + \aib \bjj X /2!+...
] +...
\]
We have assumed that $k_{0}=\bar k_{0}$.
Let
\be
Y = X + \ai \p X + \at \pp X + \al _{3} \ppp X /2! +...+
\aib \bar{\partial} X + \atb \bj X + \bar{\al _{3}} \bjj X /2! +...
\ee
Then defining $x_{n}, \bar{x_{n}}$ in the same way as before (2.6),(2.7)
\be
\frac{\p Y}{\p x_{1}} = \p X + \ai \pp X + \at \ppp X /2! +...
\ee
\be
\frac{\p Y}{\p \xb _{1}}= \bar {\p} X + \aib \bj X + \atb \bjj X /2! +...
\ee
\be
\frac{\pp Y}{\p x_{1} \p \xb _{1}} = 0
\ee
The exponent becomes
\be
\ko Y +  \sum _{n} ( k_{n}
\frac{\p Y}{\p x_{n}} +
\bar{k_{n}}
\frac{\p Y}{\p \xb _{n}})
\ee
As before we can define $<YY> =\sigma  $  to get
\be
<\frac{\p Y}{\p x_{n}} Y> = 1/2 \frac{\p \s}{\p x_{n}} \, ;
 <\frac{\p Y}{\p x_{n}}  \frac{\p Y}{\p x_{m}}> = 1/2
 ( \frac{\pp \s}{\p x_{n} \p x_{m}} - \frac{\p \s}{\p x_{n+m}} ) \, ;
\ee
\[
 <\frac{\p Y}{\p x_{n}}  \frac{\p Y}{\p \xb _{m}}> = 1/2
 ( \frac{\pp \s}{\p x_{n} \p \xb _{m}} )
\]
The loop variable (5.1) along with its '$\sigma $' dependence is
\be
exp(i(\ko Y +  \sum _{n} ( k_{n}
\frac{\p Y}{\p x_{n}} +
\bar{k_{n}}
\frac{\p Y}{\p \xb _{n}}) + ( \ko ^{2} + \qo ^{2}) \s + (\ki . \ki
+\qi \qi )1/2(\dsii - \dst )
\ee
\[
+ (\ki .\ko + \qi . \qo ) \dsi +
( \kt . \ko + \qt \qo ) \dst
+ ( \kib . \kib + \qib \qib ) 1/2 ( \dsiib - \dstb )
\]
\[
+ ( \kib . \ko
 + \qib \qo ) \dsib + ( \ktb . \ko + \qtb \qo ) \dstb
+( \ki . \kib + \qi \qib ) \dsb ))
\]
We have denoted the D+1 st component of '$k$' by $q$ and integrated
out the $D+1$st component of $X$.  (We are following the procedure
of \cite{BSI} rather than the Kaluza Klein approach of Sec. II in this
paper.)  The vertex operators at the lowest mass level are \yim
\yin and $ \frac{\partial ^{2}\sigma}{\partial x_{1} \partial
\bar{x_{1}}}$.
We collect terms of this dimension and set the variation $\frac{\delta}
{\delta \sigma}$ to zero (evaluating the derivative at the point
$\frac{\partial \sigma}{\partial x_{1}} = \frac{\partial \sigma}
{\partial \bar{x_{1}}}=0$, just as in Ref.1).

There are three types of terms that need to be varied:
\be
I. \, \,
e^{(\ko ^{2} + \qo ^{2}) \s} \e \{ (i)^{2} \kim \kinb
\frac{\p Y^{\mu}}{\p x_{1}}
\frac{\p Y^{\nu}}{\p \xb _{1}}   \}
\ee

\be
II. \, \,
e^{(\ko ^{2} + \qo ^{2}) \s} \e \{
(\ki . \ko + \qi \qo ) \dsi i \kimb \frac{\p Y^{\nu}}{\p \xb _{1}}
\ee
\[
+(\kib .\ko + \qib \qo ) \dsib i \kim
\frac{\p Y^{\nu}}{\p x _{1}} \}
\]

\be
III. \, \,
e^{(\ko ^{2} + \qo ^{2}) \s} \e \{
(\ki .\ko + \qi \qo )(\kib \ko + \qib \qo ) \dsi \dsib
\ee
\[
+2(\ki . \kib + \qi \qib ) 1/2 \dsb \}
\]

\be
\dds I = -(\ko^{2} + \qo^{2} ) \kim \kinb
\frac{\p Y^{\mu}}{\p x _{1}}
\frac{\p Y^{\nu}}{\p \xb  _{1}}
\e
\ee

\be
\dds II = (\ki . \ko + \qi \qo ) \kimb \kon
\frac{\p Y^{\mu}}{\p \xb  _{1}}
\frac{\p Y^{\nu}}{\p x _{1}}
\ee
\[
+(\kib . \ko + \qib \qo ) \kim \kon
\frac{\p Y^{\nu}}{\p \xb  _{1}}
\frac{\p Y^{\mu}}{\p x _{1}}
\]

\be
\dds III = -2(\ki .\ko + \qi \qo )( \kib .\ko + \qib \qo ) \dsb
\ee
\[
+2(\ko^{2} + \qo^{2})(\ki .\kib + \qi \qib ) \dsb
\]
\[
+(i)^{2} \kom \kon
\frac{\p Y^{\nu}}{\p \xb  _{1}}
\frac{\p Y^{\mu}}{\p x _{1}}
(\ki .\kib + \qi \qib )
\]
Coefficient of
\be
\frac{\p Y^{\nu}}{\p \xb  _{1}}
\frac{\p Y^{\mu}}{\p x _{1}}
\ee
\[
[-\ko^{2} \kim \kinb + \ki .\ko \kimb \kon + \kib . \ko \kim \kon
- \kon \kom (\ki . \kib + \qi \qib ) ] =0
\]

Coefficient of
\be
\dsb
\ee
\[
[-(\ki . \ko ) (\kib . \ko ) + \ko ^{2} ( \ki . \kib + \qi \qib )] =0
\]
We have set the $(mass)^{2}=q_{0}^{2}=0$.
The fields are the graviton, the anti symmetric tensor and the dilaton:
\be
\int [dk d\bar{k} dqd\bar{q} ] k_{1}^{(\mu } \kib ^{\nu )} \Phi [k,
\bar{k} , q, \bar{q}]
= h^{\mu \nu}
\ee
\[
\int [dk d\bar{k} dqd\bar{q} ] k_{1}^{[\mu } \kib ^{\nu ]} \Phi [k,
\bar{k} , q, \bar{q}]
= A^{\mu \nu}
\]

and
\be
\int [dk d\bar{k} dqd\bar{q} ] \qi \qib \Phi [k,
\bar{k} , q, \bar{q}] = \eta
\ee

The gauge transformations are
\be
\ki \rightarrow \ki + \li \ko \, ; \, \kib \rightarrow \kib + \bar{\li}
\ko \, ; \, \qi \rightarrow \qi \, ; \, \qib \rightarrow \qib
\ee
Since $q_{0}$ is zero, $q_{1}, \bar{q_{1}}$ remain invariant.  This
gives the usual infinitesimal transformation for the graviton
\be
h^{\mu \nu}   \rightarrow
h^{\mu \nu}    +
 \p ^{(\mu} \epsilon ^{\nu )}  + \p ^{(\nu} \bar{\epsilon }^{\mu )} =
h^{\mu \nu}    + \p ^{(\mu} \epsilon _{S}^{\nu )}
\ee
where we define
\be
\epsilon _{S}^{\mu} = \epsilon ^{\mu} + \bar{\epsilon} ^{\mu}
\ee
and
\be
A^{\mu \nu} \rightarrow A^{\mu \nu } + \p ^{[ \mu } \epsilon ^{\nu ]}
+ \p ^{[ \nu } \bar{\epsilon}^{\mu ]} = A^{\mu \nu} + \p ^{[\mu}
\epsilon _{A} ^{\nu]}
\ee
where \be
\epsilon _{A}^{\mu} = \epsilon ^{\mu} - \bar{\epsilon} ^{\mu}
\ee
and
\be
\delta \eta = 0
\ee
One can easily check that under the variations (5.20) equations
(5.16) and (5.17) are invariant.  Presumably one can reproduce these
results in the Kaluza Klein approach of Sec. II.  We have not attempted
to do this here.
\newpage
\section{Conclusion}

  In this paper we have attempted to extend the results of \cite{BSI} in
several directions and also to understand a bit better the connection
between these results and the usual BRST formalism - in particular
the issue of the difference in the form of the gauge transformation.
We have been able to understand the mechanism
of the gauge invariance , the connection with the Virasoro group and
also the origin of the ad-hoc constraints in \cite{BSI}.  We have also
extended the treatment to closed strings.  In the process of trying
 to understand some of the peculiar features there arose some open
questions.  Some of the outstanding questions are :
How does one start from the usual
BRST formalism and arrive at the equations and gauge transformations
here, at the level of the string field rather than component
by component? In particular what is the connection between $\al (t)$
and the ghost? Finally we would like to generalize these results to
the interacting theory where the issue of regularization becomes important
when you try to go off shell.   These issues have been discussed
in a related context in \cite{EWII}. We hope to report on these questions
in the future.

\underline{Acknowledgements:}
I would like to thank Martin Rocek and the members of the ITP at
Stony Brook for their hospitality. I would also like to thank Warren Siegel
for some useful discussions.


\newpage
\begin{thebibliography}{99}
\bibitem{SZ} W. Siegel, Phys. Lett B142(1984)276, B151(1985)391;
W. Siegel and B. Zwiebach, Nucl. Phys. B263(1986)105.
\bibitem{EW} E. Witten, Nucl. Phys. B268(1986)253.
\bibitem{Z} B. Zwiebach, IASSNS-HEP-92/41, Jun 1992
\bibitem{BSI} B. Sathiapalan, Nucl. Phys. B326(1989)376.
\bibitem{LFT} Phys. Lett.B135 (1984)75,
Nucl. Phys. B273 (1986) 413;
E. Fradkin and A. A. Tseytlin, Phys. Lett B158 (1985) 316.
\bibitem{CSen}C. G. Callan, D. Friedan, E. Martinec and M. Perry,
Nucl. Phys. B262 (1985) 593; A. Sen Phys. Rev. D32 (1985) 2102;
C. G. Callan and Z. Gan, Nucl. Phys. B272 (1986) 647.
\bibitem{DS}S. R. Das and B. Sathiapalan,
Phys. Rev. Lett 56 (1986)2664,
57(1986)1511.
\bibitem{BSII} B. Sathiapalan, Phys. Lett. B201(1988) 454.
\bibitem{BM} T. Banks and E. Martinec,
Nucl. Phys. B294 (1987) 733.
\bibitem{HLP} J. Hughes, J. Liu and J. Polchinski,
Nucl. Phys. B316(1989)15
\bibitem{BSIII} B. Sathiapalan, Nucl. Phys. B376(1992) 387.
\bibitem{BSIV} B. Sathiapalan, PSU Preprint, Jan 1993,
to appear in Nucl. Phys. B.
\bibitem{AMP} A. M. Polyakov, 'Gauge Fields and Strings',
(Harwood Academic Publishers, New York,1987).
\bibitem{OA} O. Alvarez, Nucl. Phys. B216, 125 (1983).
\bibitem{CMNP} A. Cohen, G. Moore, P Nelson and J. Polchinski,
Nucl. Phys. B267(1986)143.
\bibitem{R}A. N. Redlich, Nucl. Phys. B304(1988) 129.
\bibitem{EWII}E. Witten, IASSNS-HEP-92/53,92/63, K. Li and E. Witten
IASSNS-HEP-93/7.
\end{thebibliography}
\end{document}